# Reconstructing a Graph from Path Traces

Vincent Gripon
Electronics Department
Télécom Bretagne
Brest, France
Email: vincent.gripon@ens-cachan.org

Michael Rabbat
Dep. of Electrical & Computer Eng.
McGill University
Montréal, Canada
Email: michael.rabbat@mcgill.ca

*Abstract*—This paper considers the problem of inferring the structure of a network from indirect observations. Each observation (a "trace") is the unordered set of nodes which are activated along a path through the network. Since a trace does not convey information about the order of nodes within the path, there are many feasible orders for each trace observed, and thus the problem of inferring the network from traces is, in general, ill-posed. We propose and analyze an algorithm which inserts edges by ordering each trace into a path according to which pairs of nodes in the path co-occur most frequently in the observations. When all traces involve exactly 3 nodes, we derive necessary and sufficient conditions for the reconstruction algorithm to exactly recover the graph. Finally, for a family of random graphs, we present expressions for reconstruction error probabilities (false discoveries and missed detections).

## I. INTRODUCTION

Topology inference problems arise in a number of settings including systems biology [1], wireless communications [2], and more generally, for finding the structure of probabilistic graphical models [3]. We consider the problem of reconstructing the topology of a network from observations of subsets of nodes which form a path in the network. Specifically, for every elementary path we observe the set of nodes in the path but not their order within the path. Given such observations, the goal is to reconstruct the network topology. Since the nodes of the network are known in advance, network reconstruction boils down to inferring which pairs of nodes are and are not connected by links. Each observation (which we refer to as a "trace" below) corresponds to a subset of nodes which can be connected by a path through the network. Such a network reconstruction problem may arise in the context of biological networks [1], telecommunication networks [4], [5], brain networks [6], social media [7], or social networks [8].

When is it possible to reconstruct a graph from such incomplete data? And when it is possible, how can one reconstruct the network? We propose an algorithm for reconstructing networks from a collection of traces. The algorithm greedily inserts edges based on which pairs of nodes co-occur in observations most frequently. When only traces containing three nodes are observed (i.e., corresponding to paths of three nodes), we provide necessary and sufficient conditions for when our algorithm is guaranteed to accurately recover the edge structure of the network. The conditions are related to the notion of "triadic closure" in the social network literature [9], [10]. Then, for a class of random graphs, we derive the probabilities that our algorithm will omit true edges and include false edges.

Our previous work [5] described an algorithm for inferring graph structure from traces. The previous algorithm assumes that the path underlying each trace is generated by taking a random walk on the graph. The previous algorithm is iterative and is guaranteed to converge to a fixed point. Although the experimental results are promising, no theoretical guarantees are available for the accuracy of the network estimates it produces. In contrast, the algorithm presented in this paper is not based on any generative model, is not iterative, and conditions are provided under which it is guaranteed to exactly reconstruct the original graph.

The paper is organized as follows. Section II introduces notation, formally states the problem, and presents two examples illustrating that the problem is non-trivial. Section III describes the proposed reconstruction algorithm. Section IV focuses on the case when only traces of length 3 are observed and provides necessary and sufficient conditions under which the algorithm of Section III is guaranteed to perfectly reconstruct the graph. Section V considers the classical Erdös–Rényi random graph model and provides expressions for error rates as a function of edge density. We conclude in Section VI.

## II. NETWORK RECONSTRUCTION FROM PATH TRACES

We consider a graph $G = (V, E)$ with vertex set $V$ and edge set $E \subset V \times V$. Throughout this paper, all graphs are simple and undirected[1], and the vertex set is enumerated by the integers $V = \{1, 2, \ldots, |V|\}$. We say that two nodes $u$ and $v$ are *neighbors* in $G$ if $(u, v) \in E$. A *path* $P$ of length $m$ in $G$ is a sequence of vertices $v_1 \leftrightarrow v_2 \leftrightarrow \ldots \leftrightarrow v_m$ such that $(v_i, v_{i+1}) \in E$ for all $i = 1, \ldots, m-1$. An *elementary path* is one where no vertex appears twice. The *trace* $T$ of an elementary path $P$ is the (unordered) set of vertices appearing in $P$; the trace $T$ thus does not contain any information about the order of these vertices in $P$. The *size* of a trace $T$ is the number of vertices in the path.

### A. Problem Statement

Given a graph $G$ and an integer $k \geq 1$, let $\mathcal{P}_k(G)$ denote the set of all elementary paths in $G$ of length $k$, and let $\mathcal{T}_k(G)$

---

[1]A simple graph is one with no self-loops $(v, v)$ or multiple edges, and an undirected or symmetric graph is one where $(u, v) = (v, u)$.

denote the corresponding set of all traces of paths in $\mathcal{P}_k(G)$. We are interested in the following problem:

*Problem:* Given $\mathcal{T}_k(G)$, recover the graph $G$.

Since one can only hope to recover the subgraph of $G$ on vertices which appear in $\mathcal{T}_k(G)$, the problem boils down to recovering the edge set $E$ from $\mathcal{T}_k(G)$. Since the vertex set will always be clear from the context, we will write $(v_1, v_2) \in G$, slightly abusing notation, to mean that the edge $(v_1, v_2)$ is in the graph $G$.

Of course, when $k = 2$, the problem is trivial since $\mathcal{T}_2(G)$ is exactly the edge set, $E$. However, in general, the problem is not solvable since there exist some graphs with exactly the same set of traces. Consider for example the complete graph $K_5$ with five nodes and $K_5$ with any edge removed.

On the other hand, there are examples of graphs which can be retrieved from their traces. The path graph on $n$ vertices is the graph $P_n$ with edge set $\{(u, u+1) : 1 \leq u \leq n-1\}$. It provides one example of a graph uniquely defined by its traces of length 3 when $n \geq 6$. Proving this result is trivial and omitted due to space constraints. It makes use of the following lemma, that is also useful in the sequel.

**Lemma 1.** *If $\{v_1, v_2, v_3\}$ is in the set of traces $\mathcal{T}_3(G)$, then at least one of the edges $(v_1, v_2)$ and $(v_1, v_3)$ is in $G$.*

## III. NETWORK RECONSTRUCTION ALGORITHM

The previous section illustrated that, under the right conditions, a graph may be recovered from its traces. In this section we describe a simple algorithm for reconstructing a graph $\widehat{G}$ from a set of traces $\mathcal{T}$. The reconstructed graph is *feasible* with respect to the trace set $\mathcal{T}$ in the sense that to each trace $T \in \mathcal{T}$, there exists a path $P$ in $\widehat{G}$ such that $T$ is the trace of $P$. The algorithm employs the notion of the *weight* $w_M(P)$ of a path $P = v_1 \leftrightarrow v_2 \leftrightarrow \ldots \leftrightarrow v_m$ with respect to a $|V| \times |V|$ matrix $M$, which is defined as $w_M(P) = \sum_{i=1}^{n-1} M(v_i, v_{i+1})$.

The reconstruction algorithm involves two passes over the set of traces.

**Stage 1:** For each entry of the $|V| \times |V|$ matrix $M$, set

$$M(u,v) = |\{\{u,v,x\} \in \mathcal{T} : x \in V\}|, \quad (1)$$

where $|A|$ denotes the cardinality of the set $A$. Of course, since traces are unordered, $M(u,v) = M(v,u)$.

**Stage 2:** Initialize $\widehat{G}$ to be the empty graph on $|V|$ nodes; i.e., the graph with no edges. Then, for each trace $T \in \mathcal{T}$, find the elementary path $P^*(T)$ which has the maximum weight $w_M(P)$ of all paths $P$ through the vertices in $T$, and add each edge in $P^*(T)$ to $\widehat{G}$; if there is a tie, then add the edges from all maximal weight paths to $\widehat{G}$.

Once edges corresponding to all traces in $\mathcal{T}$ have been added, then the algorithm returns $\widehat{G}$.

### A. Computational Complexity

The algorithm described above has complexity which is exponential in the size of the traces. Each of the stages involves making one pass over all traces $T \in \mathcal{T}$. Let $|\mathcal{T}|$ denote the number of traces in $\mathcal{T}$. If all traces are of size $|T| = k$ vertices, then the first stage has complexity $\mathcal{O}(k^2 |\mathcal{T}|)$, since for each trace $T \in \mathcal{T}$ we need to increment the counts of each pair of vertices in $T$. Also, the second stage has complexity $\mathcal{O}(k!\, k\, |\mathcal{T}|)$, since there are $k!$ orderings for each trace, computing the weight for one ordering requires $k$ operations, and there are $|\mathcal{T}|$ traces in total. Note, if all traces are of size $k$ then there are at most $|\mathcal{T}| = \mathcal{O}(\binom{|V|}{k})$ traces.

## IV. NECESSARY AND SUFFICIENT CONDITIONS FOR GRAPH RECONSTRUCTION WHEN $k = 3$

In this section we focus on the case where the goal is to reconstruct a graph $G = (V, E)$ given only the set of length-3 traces $\mathcal{T}_3(G)$. We assume that $n = |V| > 3$ and that $|E| > 2$ so the problem is non-trivial. We are interested in knowing when the algorithm described in the previous section can be guaranteed to provide an accurate reconstruction. To this end, we derive necessary and sufficient conditions for when the algorithm is guaranteed to correctly identify the presence or absence of an edge between two vertices. Since traces of length 3 only provide information about the graph structure a few hops away from each edge, it is natural that these conditions only depend on the local structure of the graph. Throughout this section, we assume the graph $G$ is fixed. Let $\widehat{G}$ denote the output of the reconstruction algorithm given input $\mathcal{T}_3(G)$, and let $M$ denote the matrix of co-occurrence counts used by the algorithm. To simplify the notation, we will suppress the dependence on $G$ and simply write $\mathcal{T}_3$ to denote the set of traces of all length-3 elementary paths in $G$.

### A. Preliminaries

The reconstruction algorithm uses the values in $M$ to determine which is the maximum-weight path for each trace, and thus $M$ governs which edges do and do not appear in $\widehat{G}$. To begin, we introduce two lemmas which characterize relationships between the co-occurrence count matrix $M$, the edge set of $G$, and the estimated edge set of $\widehat{G}$.

**Lemma 2.** *The edge $(v_1, v_2)$ is in $\widehat{G}$ if and only if there exists a vertex $u$ with $\{v_1, v_2, u\} \in \mathcal{T}_3$ and at least one of the following conditions holds:*

  i) $M(v_1, v_2) \geq M(v_1, u)$,
  ii) $M(v_1, v_2) \geq M(v_2, u)$.

**Lemma 3.** *If $(v_1, v_2)$ is in $G$ and there exists another vertex $u \neq v_1, v_2$ such that $(v_1, u) \in G$ and $(v_2, u) \notin G$, then the edge $(v_1, v_2)$ is in $\widehat{G}$.*

Detailed proofs of Lemmas 2 and 3 are omitted due to space constraints and will be provided in an extended version of the paper. Fig. 1 illustrates the idea in the proof of Lemma 3, and the proof of Lemma 2 follows from similar arguments.

### B. Main Results

Lemma 3 says that for each edge $(v_1, v_2)$ in $G$, if at least one of $v_1, v_2$ has a unique neighbor, then the traces arising from the unique neighbor will be sufficient to ensure the correct edge appears in the network. We formalize this property in the definition below.

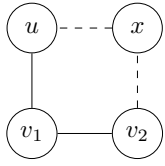

Fig. 1. Illustrating the concept in the proof of Lemma 3. By assumption, there is an edge between $v_1$ and $v_2$, and $u$ is a neighbor of $v_1$ but not $v_2$. For any trace $\{u, v_2, x\}$ containing both $u$ and $v_2$, it follows that $x$ is a neighbor of both $u$ and $v_2$ (since $u$ and $v_2$ are not neighbors). Thus, there must also be a trace $\{v_1, v_2, u\}$, and so $M(v_1, v_2) \geq M(v_2, u)$. Consequently, the edge $(v_1, v_2)$ appears in the reconstructed graph, $\widehat{G}$.

**Definition 1.** An edge $(v_1, v_2) \in G$ has the *unique neighbor* property if there exists a node $u$ such that either $(v_1, u) \in G$ and $(v_2, u) \notin G$, or $(v_2, u) \in G$ and $(v_1, u) \notin G$.

While this condition is sufficient, it is not necessary to guarantee that $(v_1, v_2)$ is in the reconstructed graph $\widehat{G}$. To obtain a necessary set of conditions we need to introduce one additional property.

**Definition 2.** An edge $(v_1, v_2) \in G$ has the *strong triadic closure* property if there exists a vertex $z$ which is neighbors of both $v_1$ and $v_2$, and any neighbor of $z$ is also a neighbor of at least one of $v_1$, and $v_2$; i.e., there exists $z \in V$ such that $(v_1, z) \in G$, $(v_2, z) \in G$, and for all $y \in V$ with $(y, z) \in G$, it also holds that either (i) $(v_1, y) \in G$, (ii) $(v_2, y) \in G$, or (iii) both $(v_1, y) \in G$ and $(v_2, y) \in G$.

Triadic closure arises in the study of social networks [9], [10], expressing the notion that "one's friends tend to also be friends". We refer to Definition 2 as the "strong" triadic closure property since it says that if $v_1$ and $v_2$ have a neighbor $z$ in common, then all other neighbors of $z$ should also be neighbors of either $v_1$ or $v_2$. In this sense, no node ends up being a hub; they are clustered together into a community of nodes which all have roughly the same number of neighbors. Taken together, the unique neighbor property and the strong triadic closure property form necessary and sufficient conditions for an edge to correctly appear in the reconstructed graph $\widehat{G}$.

**Theorem 1.** *Suppose that $(v_1, v_2) \in G$. Then $(v_1, v_2) \in \widehat{G}$ if and only if at least one of the following holds:*
  i) $(v_1, v_2)$ *satisfies the unique neighbor property;*
  ii) $(v_1, v_2)$ *satisfies the strong triadic closure property.*

   *Proof:* Suppose that $(v_1, v_2) \in G$. If $(v_1, v_2)$ satisfies the unique neighbor property, then Lemma 3 already ensures that $(v_1, v_2) \in \widehat{G}$.

Suppose, next, that $(v_1, v_2) \in G$ and $(v_1, v_2)$ satisfies the strong triadic closure property in $G$. Recall from (1) that $M(v_1, v_2)$ is equal to the number of traces of the form $\{v_1, v_2, u\}$ for some $u \in V \setminus \{v_1, v_2\}$. By the strong triadic closure property, there exists a vertex $z \in V$ which is neighbors of both $v_1$ and $v_2$, and hence the trace $\{z, v_1, v_2\}$ is in $\mathcal{T}_3$. By Lemma 1, for every other trace of the form $\{z, v_1, u\} \in \mathcal{T}_3$ with $u \neq v_2$, either $(u, v_1) \in G$ or $(u, z) \in G$. If $(u, v_1) \in G$ then the path $u \leftrightarrow v_1 \leftrightarrow v_2$ is in $G$, and so the trace $\{u, v_1, v_2\} \in \mathcal{T}_3$. If $(u, z) \in G$ and $(u, v_1) \notin G$, then by the strong triadic closure property, we must have that $(u, v_2) \in G$. In this case, the path $u \leftrightarrow v_2 \leftrightarrow v_1$ is in $G$, and so the trace $\{u, v_1, v_2\}$ is in $\mathcal{T}_3$. It follows that $M(v_1, v_2) \geq M(v_1, z)$, and by Lemma 2 we have that $(v_1, v_2) \in \widehat{G}$.

Now, for the converse, suppose that $(v_1, v_2) \in \widehat{G}$, and assume, for the sake of a contradiction, that $(v_1, v_2)$ satisfies neither the unique neighbor property nor the strong triadic closure property. Since the unique neighbor property does not hold, if $(u, v_1) \in G$ then $(u, v_2) \in G$ also. Since the strong triadic closure property does not hold, for all common neighbors $u$ of $v_1$ and $v_2$, there exists another vertex $x_u \neq v_1, v_2$ which is a neighbor of neither $v_1$ nor $v_2$; i.e., $(x_u, v_1) \notin G$ and $(x_u, v_2) \notin G$. The reconstruction algorithm only inserts an edge $(v_1, v_2)$ into $\widehat{G}$ if there is at least one trace of the form $\{v_1, v_2, u\} \in \mathcal{T}_3$. For each vertex $u$ such that $\{v_1, v_2, u\} \in \mathcal{T}_3$, since the unique neighbor property does not hold, it must be true that $(v_1, u) \in G$ and $(v_2, u) \in G$. Moreover, since the strong triadic closure property does not hold, there must be another vertex $x_u$ with $(u, x_u) \in G$ and for which $(v_1, x_u) \notin G$ and $(v_2, x_u) \notin G$. It follows that the paths $v_1 \leftrightarrow u \leftrightarrow x_u$ and $v_2 \leftrightarrow u \leftrightarrow x_u$ are in $G$, and so traces $\{v_1, u, x_u\}$ and $\{v_2, u, x_u\}$ are in $\mathcal{T}_3$. Thus, for every trace $\{v_1, v_2, u\} \in \mathcal{T}_3$, there are also traces $\{v_1, u, x_u\} \in \mathcal{T}_3$ and $\{v_2, u, x_u\} \in \mathcal{T}_3$. Consequently, $M(v_1, v_2) < M(v_1, u)$ and $M(v_1, v_2) < M(v_2, u)$. Then, by Lemma 2, we conclude that $(v_1, v_2) \notin \widehat{G}$, which is a contradiction. Thus, if $(v_1, v_2) \in \widehat{G}$, then $(v_1, v_2)$ must satisfy the unique neighbor property or the strong triadic closure property. ∎

Theorem 1 provides necessary and sufficient conditions under which the reconstruction algorithm will return a network $\widehat{G}$ that contains all edges which are in $G$ (i.e., no missed detections). At the same time, we would like to ensure that $\widehat{G}$ contains no edges which are not in $G$ (i.e., no false positives). To this end we introduce the following two properties regarding the graph structure around pairs of nodes that are not neighbors.

**Definition 3.** A pair of vertices $v_1$ and $v_2$ for which $(v_1, v_2) \notin G$ has the *distinct neighbors* property if there exist a vertex $v_1'$ which is a neighbor of $v_1$ and not a neighbor of $v_2$, and there exists a vertex $v_2'$ which is a neighbor of $v_2$ and not a neighbor of $v_1$.

**Definition 4.** A pair of vertices $v_1$ and $v_2$ for which $(v_1, v_2) \notin G$ has the *weak triadic closure* property if for every vertex $u$ which is a neighbor of both $v_1$ and $v_2$, there exists another vertex $x_u$ which is neighbors with $u$ and is not neighbors with both $v_1$ and $v_2$; i.e., for every $u \in V$ with $(u, v_1) \in G$ and $(u, v_2) \in G$, there exists $x_u \in V$ such that $(u, x_u) \in G$ and either:
  i) $(v_1, x_u) \in G$ and $(v_2, x_u) \notin G$, or
  ii) $(v_2, x_u) \in G$ and $(v_1, x_u) \notin G$, or
  iii) $(v_1, x_u) \notin G$ and $(v_2, x_u) \notin G$.

The distinct neighbors and weak triadic closure properties form a set of necessary and sufficient conditions to guarantee that an edge is not falsely inserted between a pair of nodes in the reconstructed graph.

**Theorem 2.** *Suppose that $(v_1, v_2) \notin G$. Then $(v_1, v_2) \notin \widehat{G}$ if and only if at least one of the following holds:*
  i) $v_1$ *and* $v_2$ *has the distinct neighbors property;*
  ii) $(v_1, v_2)$ *has the weak triadic closure property.*

*Proof:* Suppose that $(v_1, v_2) \notin G$. Let $U = \{u \in V : \{v_1, v_2, u\} \in \mathcal{T}_3\}$ denote the set of vertices which co-occur in some trace with $v_1$ and $v_2$. By definition, $M(v_1, v_2) = |U|$, and $(v_1, v_2) \notin \widehat{G}$ if $M(v_1, v_2) = 0$, since the reconstruction algorithm only inserts edges between pairs of nodes which co-occur in at least one trace. Thus, for the remainder of the proof we suppose that $M(v_1, v_2) > 0$. Then for each $u \in U$, by Lemma 1, it must be true that $(v_1, u) \in G$ and $(v_2, u) \in G$. If $M(v_1, v_2) > 1$, then there exist $u_1, u_2 \in U$, $u_1 \neq u_2$, and both of the paths $u_1 \leftrightarrow v_1 \leftrightarrow u_2$ and $u_1 \leftrightarrow v_2 \leftrightarrow u_2$ are in $G$. Consequently, the traces $\{u_1, u_2, v_1\}$ and $\{u_1, u_2, v_2\}$ are both in $\mathcal{T}_3$.

Suppose that $v_1$ and $v_2$ have the distinct neighbors property, and let $v_1'$ denote the distinct neighbor of $v_1$ and $v_2'$ denote the distinct neighbor of $v_2$. Fix a $u' \in U$, and for $i = 1, 2$ define

$$\begin{aligned}\mathcal{S}_i &= \{\{u', v_i, u\} \in \mathcal{T}_3 : u \in U, u \neq u'\} \\ &\cup \{u', v_1, v_2\} \cup \{u', v_i, v_i'\}.\end{aligned} \quad (2)$$

Thus, $\mathcal{S}_1$ (resp. $\mathcal{S}_2$) contains a subset of the traces that involve both $v_1$ and $u'$ (resp. $v_2$ and $u'$). Moreover, $|\mathcal{S}_1| > |U|$ and $|\mathcal{S}_2| > |U|$ since $\mathcal{S}_1$ (resp. $\mathcal{S}_2$) contains one trace for each $u \in U$, as well as the trace $\{u', v_1, v_1'\}$ (resp. $\{u', v_2, v_2'\}$). It follows that,

$$M(v_1, u') \geq |\mathcal{S}_1| > |U| = M(v_1, v_2), \quad \text{and} \quad (3)$$
$$M(v_2, u') \geq |\mathcal{S}_2| > |U| = M(v_1, v_2), \quad (4)$$

and therefore, by Lemma 2, $(v_1, v_2) \notin \widehat{G}$.

Suppose, instead, that $v_1$ and $v_2$ have the weak triadic closure property. Let $\tilde{u} \in U$ be a neighbor of both $v_1$ and $v_2$, and let $x_{\tilde{u}}$ denote a neighbor of $\tilde{u}$ which is not neighbors with both $v_1$ and $v_2$. Similar to above, for $i = 1, 2$ define

$$\begin{aligned}\widetilde{\mathcal{S}}_i &= \{\{\tilde{u}, v_i, u\} \in \mathcal{T}_3 : u \in U, u \neq \tilde{u}\} \\ &\cup \{\tilde{u}, v_1, v_2\} \cup \{\tilde{u}, v_i, x_{\tilde{u}}\}.\end{aligned} \quad (5)$$

Since $x_{\tilde{u}}$ is not neighbors with both $v_1$ and $v_2$, and since $v_1$ and $v_2$ are not neighbors, we have $x_{\tilde{u}} \notin U$. Consequently, $|\widetilde{\mathcal{S}}_i| > |U|$, for $i = 1, 2$, and so, similar to above,

$$M(v_1, \tilde{u}) \geq |\widetilde{\mathcal{S}}_1| > |U| = M(v_1, v_2), \quad \text{and} \quad (6)$$
$$M(v_2, \tilde{u}) \geq |\widetilde{\mathcal{S}}_2| > |U| = M(v_1, v_2), \quad (7)$$

and so $(v_1, v_2) \notin \widehat{G}$ by Lemma 2.

Now, for the converse, suppose that $(v_1, v_2) \notin G$ and $(v_1, v_2) \notin \widehat{G}$. We would like to show that $v_1$ and $v_2$ must have either the distinct neighbors property or the star property (or both). Suppose, for the sake of a contradiction, that vertices $v_1$ and $v_2$ have neither the distinct neighbors property nor the star property. Since $v_1$ and $v_2$ do not have the distinct neighbors property, either every neighbor of $v_1$ is also a neighbor of $v_2$, or every neighbor of $v_2$ is also a neighbor of $v_2$. Without loss of generality, suppose that all neighbors of $v_1$ are also neighbors of $v_2$ (i.e., $(u, v_1) \in G \Rightarrow (u, v_2) \in G$). Since $v_1$ and $v_2$ do not have the weak triadic closure property, there exists a vertex $u \in V$ which is a neighbor of both $v_1$ and $v_2$, and all neighbors of $u$ are also neighbors of both $v_1$ and $v_2$. Consider all traces of the form $\{v_1, u, x\} \in \mathcal{T}_3$, for $x \in V$. By Lemma 1, either $(x, v_1) \in G$ or $(x, u) \in G$. If $(x, v_1) \in G$ then $(x, v_2) \in G$ also, since the distinct neighbor property does not hold. On the other hand, if $(x, u) \in G$, then $(x, v_1)$ and $(x, v_2)$ must also be in $G$, since $u$ is neighbors with both $v_1$ and $v_2$ and they do not have the weak triadic closure property. Since for every trace $\{v_1, u, x\}$, there is a trace $\{x, v_1, v_2\}$, it follows that $M(v_1, u) \leq M(v_1, v_2)$, and by Lemma 2 we conclude that $(v_1, v_2) \in \widehat{G}$, which is a contradiction. Therefore, if $(v_1, v_2) \notin \widehat{G}$ the $v_1$ and $v_2$ have the distinct neighbors property or the weak triadic closure property. ∎

## V. Reconstruction Errors in Random Graphs

Next we suppose that the graph $G$ follows the well-known Erdös–Rényi [11], [12] model for random graphs, and we provide expressions for the edge false alarm and missed detection probabilities based on the properties defined in the previous section. The Erdös–Rényi model, $\mathcal{G}_{n,p}$, is a random graph on $n = |V|$ nodes, where each possible edge $(u, v) \in V \times V$ is present with probability $p$, independent of all other edges. Let $\mathbb{P}(\cdot)$ denote the corresponding probability distribution over graphs; we suppress the dependence on $n$ and $p$ to simply the notation. As in the previous section, let $\mathcal{T}_3$ denote the set of traces of length-3 elementary paths in $G$, and let $\widehat{G}$ denote the graph obtained by running the reconstruction algorithm on $\mathcal{T}_3$.

### A. Edge Missed Detection Probability

First, we provide an expression for the probability that the reconstruction algorithm misses an edge in $\widehat{G}$. Condition on the event that $(v_1, v_2) \in G$. From Theorem 1 the event $(v_1, v_2) \notin \widehat{G}$ occurs if the edge has neither the unique neighbor property nor the strong triadic closure property. Let

$$U = \{u \in V : (u, v_1) \notin G \text{ and } (u, v_2) \notin G\}$$

denote the set of vertices which are not neighbors with either $v_1$ or $v_2$, and let

$$Z = \{z \in V : (z, v_1) \in G \text{ and } (z, v_2) \in G\}$$

denote the set of vertices which are neighbors of both $v_1$ and $v_2$. If the unique neighbor property does not hold then the sets $U$ and $Z$ partition the vertices $V \setminus \{v_1, v_2\}$. Conditional on the event that $(v_1, v_2) \in G$, this occurs with probability

$$\sum_{k=0}^{n-2} \binom{n-2}{k} (1-p)^{2k} p^{2(n-2-k)}. \quad (8)$$

Next, if $(v_1, v_2)$ does not have the strong triadic closure property, then there exists a node $z \in Z$ which is neighbors of both $v_1$ and $v_2$, and $z$ has a neighbor $u \in U$. Then, conditional on $(v_1, v_2) \in G$, the unique neighbor property not holding, and $|U| = k$, this occurs with probability

$$(1 - (1-p)^k)^{n-2-k}. \qquad (9)$$

Putting these together, we have that

$$\mathbb{P}\big((v_1, v_2) \notin \widehat{G} \mid (v_1, v_2) \in G\big) \qquad (10)$$
$$= \sum_{k=0}^{n-2} \binom{n-2}{k}(1-p)^{2k}(p^2(1-(1-p)^k))^{n-2-k}.$$

### B. Edge False Alarm Probability

Next, we provide an expression for the probability that the reconstruction algorithm erroneously adds an edge in $\widehat{G}$. From Theorem 2, we know that $(v_1, v_2) \in \widehat{G}$ only if neither the distinct neighbors property nor the weak triadic closure property hold. Let

$$U_1 = \{u \in V : (u, v_1) \in G \text{ and } (u, v_2) \notin G\}$$

denote the set of nodes which are only neighbors of $v_1$, and, similarly, let $U_2$ denote the set of node which are only neighbors of $v_2$. If the distinct neighbors property does not hold, then at least one of $U_1$ and $U_2$ are empty. Suppose that $(v_1, v_2) \notin G$ and $|Z| = k$, where $U$ and $Z$ are as above. The probability that at least one of $U_1$ and $U_2$ is empty, in which case the distinct neighbors property does not hold, is

$$A(n, p, k) = 1 - (1 - (1-p)^{n-2-k})^2.$$

Next, for the weak triadic neighbors not to hold, for every vertex $z \in Z$, any neighbors of $z$ must also be in $Z$. Again, condition on the events that $|Z| = k$ and that $(v_1, v_2) \notin G$. Fix a node $z \in Z$. The probability that no $z \in Z$ is connected to any node in $V \setminus (Z \cup \{v_1, v_2\})$, in which case the weak triadic closure property odes not hold, is

$$B(n, p, k) = 1 - (1 - (1-p)^{n-2-k})^k.$$

Putting these together, we have that

$$\mathbb{P}\big((v_1, v_2) \in \widehat{G} \mid (v_1, v_2) \notin G\big)$$
$$= \sum_{k=0}^{n-2} \binom{n-2}{k} p^{2k} A(n, p, k) B(n, p, k). \qquad (11)$$

### C. Edge Error Rate

An expression for the edge error rate is obtained using the false alarm and missed detection probabilities derived above. We validate this edge error rate expression via simulation. Figure 2 shows three curves, corresponding to edge probabilities $p = 0.1$, $0.8$, and $0.5$ (from top to bottom) in the Erdös–Rényi model, for networks of up to $n = 100$ nodes. The curve shows the theoretical error rate, and the symbols indicate the empirical edge error rates found by Monte Carlo simulation.

It is interesting to note that for any fixed $p$, this error rate goes to 0 as $n$ tends to infinity.

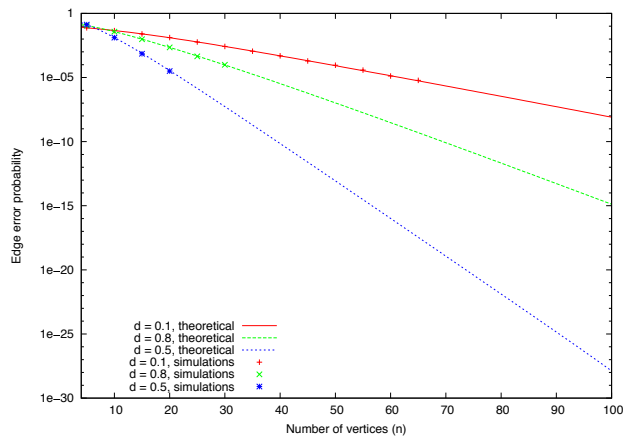

Fig. 2. Results of numerical experiments validating the edge error rate expression.

## VI. DISCUSSION

The results presented here focus on recovery of networks from all traces of length 3. In the future it would be interesting to extend these results to situations where longer traces are available, where only a subset of traces is observed, and/or where the traces are of heterogeneous sizes. It is also of interest to characterize the smallest number of traces required to accurately recover network structure. Likely, this depends on characteristics of the network structure itself.

### REFERENCES

[1] L. Acharya, T. Judeh, G. Wang, and D. Zhu, "Optimal structural inference of signaling pathways from overlapping and unordered gene sets," *Bioinformatics*, vol. 28, no. 4, pp. 546–556, 2012.
[2] J. Yang, S. Draper, and R. Nowak, "Passive learning of the interference graph of a wireless network," in *Proc. IEEE ISIT*, Boston, USA, Jul. 2012.
[3] A. Das, P. Netrapalli, S. Sanghavi, and S. Vishwanath, "Learning Markov graphs up to edit distance," in *Proc. IEEE ISIT*, Boston, USA, Jul. 2012.
[4] M. Rabbat, J. Treichler, S. Wood, and M. Larimore, "Understanding the topology of a telephone network via internally-sensed network tomography," in *Proc. IEEE ICASSP*, Philadelphia, PA, Mar. 2005.
[5] M. Rabbat, M. Figueiredo, and R. Nowak, "Network inference from co-occurrences," *IEEE Trans. Info Theory*, vol. 54, no. 9, pp. 4053–4068, Sep. 2008.
[6] P. Hagmann, L. Cammoun, X. Gigandet, R. Meuli, C. Honey, V. Weeden, and O. Sporns, "Mapping the structural core of human cerebral cortex," *PLoS biology*, vol. 6, no. 8, p. e159, Jul. 2008.
[7] M. Gomez-Rodriguez, J. Leskovec, and A. Krause, "Inferring networks of diffusion and influence," in *Proc. ACM KDD*, 2010.
[8] J. Silva and R. Willett, "Hypergraph-based detection of anomalous high-dimensional co-occurrences," *IEEE Trans. Pattern Analysis and Machine Intelligence*, vol. 31, no. 3, pp. 563–569, Mar. 2009.
[9] M. Granovetter, "The strength of weak ties," *American Journal of Sociology*, vol. 78, no. 6, pp. 1360–80, May 1973.
[10] D. Easley and J. Kleinberg, *Networks, Crowds, and Markets: Reasoning About a Highly Connected World*. Cambridge University Press, 2010.
[11] P. Erdös and A. Rényi, "On random graphs I." *Publicationes Mathematicae*, vol. 6, pp. 290–297, 1959.
[12] B. Bollobás, *Random Graphs*, 2nd ed. Cambridge University Press, 2001.